\documentclass[journal=jacsat,manuscript=article]{achemso}

\usepackage[version=3]{mhchem} 
\usepackage{float}

\author{Amalya C. Johnson}
\affiliation[Stanford University]
{Department of Materials Science \& Engineering, Stanford University, Stanford, CA, 94305, USA}
\author{Sorren Warkander}
\affiliation[Lawrence Berkeley National Laboratory]
{Chemical Sciences Division, Lawrence Berkeley National Laboratory, Berkeley, CA, 94720, USA}
\author{Archana Raja}
\affiliation[Lawrence Berkeley National Laboratory]
{Molecular Foundry, Lawrence Berkeley National Laboratory, Berkeley, CA, 94720, USA}
\alsoaffiliation[Kavli Energy NanoScience Institute]
{Kavli Energy NanoScience Institute, University of California Berkeley, Berkeley, CA 94720, USA}
\author{Fang Liu}
\affiliation[Stanford University]
{Department of Chemistry, Stanford University, Stanford, CA, 94305, USA}
\email{fliu10@stanford.edu}

\title[An \textsf{achemso} demo]
  {Extreme Thermal Insulation in Nano-Bubble Wrap Materials}

\abbreviations{TDTR, 2D Materials, TMDCs}

\begin{document}


\begin{abstract}
Achieving ultra-low thermal conductivity under ambient conditions is a fundamental challenge constrained by classical heat transport limits and material design trade-offs. Here, we introduce a new class of nano-bubble wrap architectures that achieve exceptionally low thermal conductivity by integrating nanoscale gas confinement with atomically thin, weakly coupled van der Waals solids. Using scalable patterning of 2D monolayers into periodic nano-bubbles and nano-wrinkles, we construct materials with structural analogies to macroscopic bubble wrap but engineered at length scales much shorter than the mean free path of air and the mean free path of phonons in the atomically thin monolayers. Time-domain thermoreflectance measurements reveal out-of-plane thermal conductivities nearly an order of magnitude lower than that of air and commercial aerogels, reaching critical values below 0.001 W $\cdot$ M$^{-1}$K$^{-1}$ under room temperature and atmospheric pressure. This extreme thermal resistance arises from the combined suppression of gas-phase conduction, phonon transport, and interfacial coupling. Our findings establish nano-bubble wraps as a versatile platform for tuning heat flow in ultrathin materials and open new pathways for designing thermal metamaterials and energy-efficient technologies.
\end{abstract}

\section{Introduction}

Thermal transport is a fundamental topic in the physical sciences, with both extremes of thermal conductivity playing crucial roles in the energy efficiency and thermal management of nanoscale devices to large-scale infrastructure. On the insulating end, many conventional materials limit heat flow by trapping air within porous frameworks, as seen in fibrous materials like wool or cotton. When the characteristic pore size approaches or falls below the mean free path of gas molecules ($\approx$ 69 nm for air at ambient conditions \cite{fu_critical_2022}), thermal transport enters the Knudsen regime, where classical macroscopic laws of heat transport break down, and thermal conductivity can be significantly suppressed. A well-established example of this principle is aerogel, which has long held the record for the lowest thermal conductive material over decades. Aerogels are composed of interconnected organic or inorganic networks enclosing pores typically tens to hundreds of nanometers in size, achieving thermal conductivities close to or below that of air \cite{fu_critical_2022}. However, further suppression of thermal transport in aerogels is fundamentally limited by two intrinsic factors: open porosity, which permits residual gas conduction and radiative transfer, and covalently bonded solid frameworks that allow phonon transport. An ideal strategy to overcome these limitations involves the use of atomically thin, van der Waals–coupled solid frameworks that enclose fully confined nanoscale gas pockets. When the dimensions of these gas domains fall below both the molecular and phonon mean free paths, thermal transport through both gas-phase collisions and lattice vibrations can be effectively minimized—enabling a new class of ultralow thermal conductivity materials. 

Here, we introduce a new class of materials to implement this design principle: nano-bubble wrap s. Inspired by conventional bubble wrap structure, which isolates mechanical vibrations via air-filled cavities embedded in flexible films \cite{chavannes_method_1964}, its nanoscale analogue enables spatial modulation of mechanical and thermal coupling with nanometer precision. This architecture is made possible by the advent of monolayer two-dimensional (2D) materials, which combine atomically thin geometries with intrinsically anisotropic phonon dynamics. \cite{sood_engineering_2021, sood_quasi} The strong in-plane covalent bonds of monolayers lead to moderate\cite{hong_thermal_2016, liu_measurement_2014} to high \cite{seol_two-dimensional_2010, chen_raman_2011} in-plane thermal conductivities. However, this in-plane conductivity is highly sensitive to lateral size and decreases substantially when flake dimensions approach the phonon mean free path, due to surface scattering \cite{zhou_first-principles_2015}. On the other hand, the weak interlayer coupling in layered vdW solids significantly impedes cross-plane phonon transport. The cross-plane interfacial thermal conductance can be tuned by structural modifications such as thickness-dependent phonon scattering \cite{sood_engineering_2021, sood_quasi}, rotational mismatch \cite{kim_extremely_2021}, and mass-density mismatch \cite{vaziri_ultrahigh_2019}. Together, the strong anisotropy in thermal transport and exceptional mechanical flexibility of 2D monolayers provide a powerful platform for nanoscale engineering of heat flow and the creation of materials with ultralow thermal conductivity. 

The investigation of the formation and nature of bubbles in monolayers was originally motivated by the need to mitigate interfacial contaminations. Micrometer-scale bubbles frequently arise stochastically during monolayer transfer or stacking processes, where gas, liquid, or solid adsorbates become trapped between a 2D material and its supporting substrate. Beyond spontaneous formation, a range of strategies have been developed to purposefully generate individual bubbles, including molecular intercalation through defects\cite{gasparutti_how_2020}, localized laser heating\cite{zhang_construction_2020}, voltage-induced hydrolysis \cite{an_graphene_2017}, and ion irradiation to activate trapped species\cite{stolyarova_observation_2009}. These approaches allow for controlled encapsulation of different chemical environments within bubbles. Although bubbles in 2D materials were initially regarded as defects that perturb electronic properties and quantum phases, they are increasingly recognized as functional nanoscale architectures, serving as platforms for investigating nanoconfined matter\cite{algara-siller_square_2015}, interfacial mechanics\cite{wang_measuring_2017}, strain-induced effects and emergence of single photon emitters \cite{sanchez_2d_2021}. To date, most bubbles generated in monolayers fall in the sub-micrometer to micrometer range, with dimensions largely dictated by the volume of encapsulated species.  

Here, we demonstrate a scalable strategy to engineer nanoscale structurally ordered “bubble wrap” architectures composed of periodic arrays of 1D (“nano-wrinkles”) or 0D (“nano-bubbles”) deformations of monolayers over extended areas. These intrinsic deformations enable spatial modulation of strain distribution, bandgaps, and interfacial contact. Using time-domain thermoreflectance (TDTR), we measure the thermal conductivity of various 2D nano-bubble wrap  materials and heterostructures, revealing extremely low thermal conductivity strongly modulated by the geometry. This method is highly scalable and tunable, offering new opportunities for tailoring nanoscale thermal transport. 

\section{Methods}

\begin{figure}[H]
   \centering
   \includegraphics[width=1\linewidth]{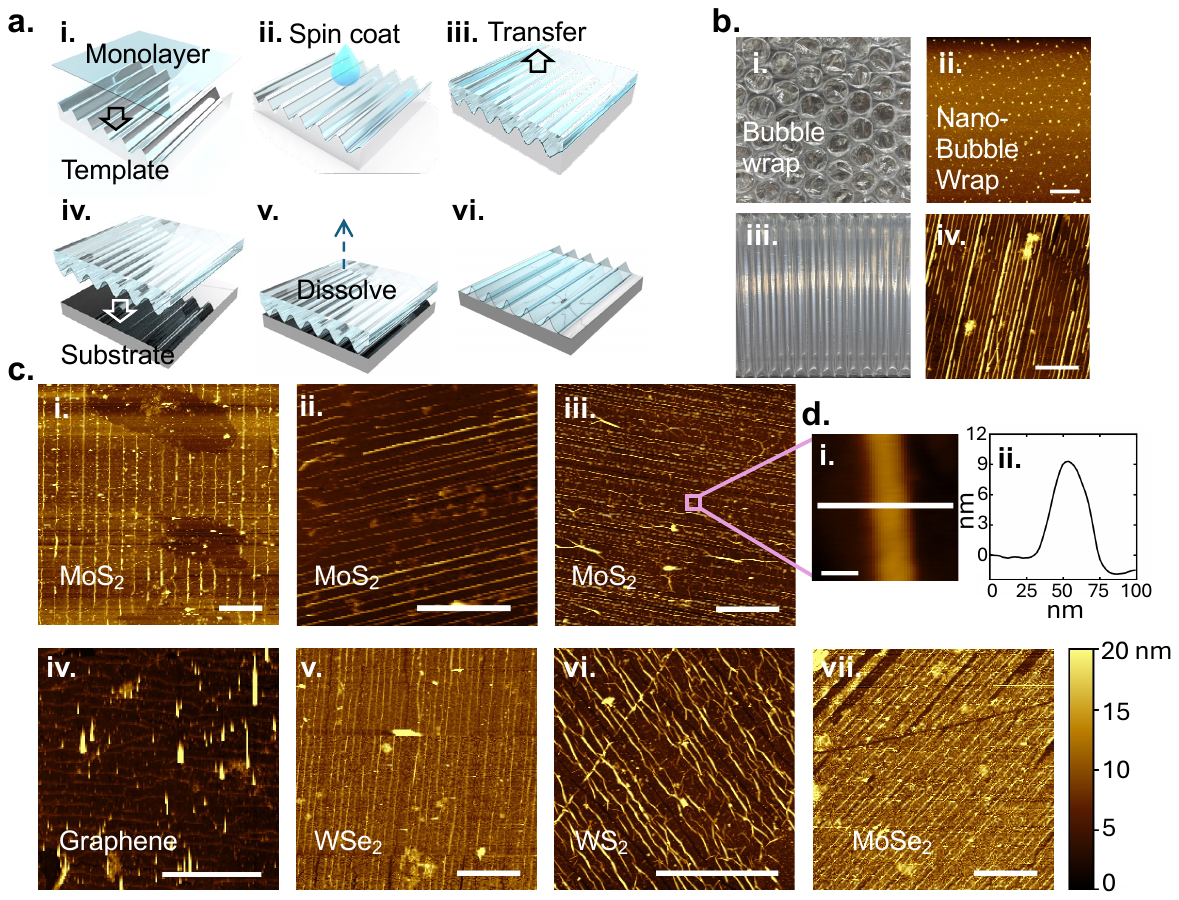}
   \caption{(a) Schematic of the fabrication process for monolayer nano-bubble wrap structures. (i) Large-area monolayers, prepared via Au-tape exfoliation, are transferred onto a rigid nanoscale-patterned template using a thin polymer support layer. (ii) A second polymer layer is spin-coated on top to conformally preserve the shape of the underlying template. (iii-iv) The polymer/monolayer stack is transferred onto the desired target substrate. (v) The polymer layers are dissolved, and (vi) the nano-deformed monolayer remains, maintaining template-induced periodic structure. (b) (i)(iii) pictures of 0D and 1D macroscopic bubble wraps and (ii)(iv) AFM images of 0D and 1D nanoscopic bubble wraps. Scale bars 5 $\mu$m. (c) AFM images of nano-deformed monolayers. (i–iii) Nano-wrinkles in MoS$_2$ with periodicities of $\approx$ 1700 nm, 530 nm, and 300 nm. (iv–vii) Nano-wrinkles in graphene, WSe$_2$, WS$_2$, and MoSe$_2$, respectively. Scale bars 5 $\mu$m. (d) (i) AFM image of an individual nano-wrinkle. Scale bar: 200 nm. (ii) Height profile extracted along the solid line in (i).} 
   \label{fig:1}
\end{figure}

A schematic of the fabrication process is shown in Figure 1a. Large-area, single-crystal monolayers are first prepared on SiO$_2$/Si substrates using gold-tape exfoliation, yielding uniform, high-quality flakes suitable for transfer\cite{liu_disassembling_2020}.  The monolayer is then spin-coated with a layer of cellulose acetate butyrate (CAB) polymer ($\approx$ 30 mg/mL in ethyl acetate, Sigma Aldrich) at 1000 rpm. The coated film is subsequently lifted off from the SiO$_2$/Si substrate using a standard water-assisted wet transfer technique\cite{schneider_wedging_2010} and transferred onto a rigid, periodically structured template. The templates are either produced with lithography or using commercial optical diffraction gratings with varying periodicities (e.g., Thorlabs GH13-36U) to define linear nanoscale topographies. After conformal contact with the template, a second layer of polymer is spin-coated at 400 rpm onto the exposed polymer surface to preserve the imprinted nanoscale deformations. The entire polymer/monolayer stack is then lifted from the template, maintaining the deformed topography of the monolayer. This composite is subsequently transferred onto a destination substrate, such as SiO$_2$/Si. Upon removal of the polymer support layers with ethyl acetate, the monolayer relaxes into a periodic, nano-bubbled or nano-wrinkled morphology with periodicity of the original template (Fig. \ref{fig:1}a, v–vi).

Figure 1b and 1c show atomic force microscopy (AFM) images of the resulting nano-deformed monolayers of WS$_2$, WSe$_2$, MoS$_2$, MoSe$_2$, and graphene. By tuning the periodicity and dimensions of the rigid template, we achieve deterministic control over the wrinkle-wrinkle spacing ($\text{d}_\text{ww}$) and geometry of the bubbles and wrinkles, as demonstrated in Figure 1c, ii–iv. Figure \ref{fig:1}d presents the AFM morphology of a representative individual nano-wrinkle, with height of $\approx$ 10 nm and FWHM of $\approx$ 25 nm. A statistical distribution of wrinkle aspect ratios is shown in Supplementary Figure S1. Despite variation in wrinkle density across different pattern geometries, the height and shape of individual wrinkles remain remarkably consistent. This uniformity likely arises from the spontaneous mechanical relaxation of the monolayer into a stable, energetically favorable configuration on a flat substrate\cite{khestanova_universal_2016}. In contrast to conventional 2D material processing, where micrometer-scale bubbles or wrinkles typically form stochastically during transfer or stacking, this method enables the reproducible fabrication of well-defined, nanometer-scale deformation arrays at designed periodicities.

\section{Spectroscopic Characterization}

\begin{figure}[H]
   \centering
   \includegraphics[width=0.9\linewidth]{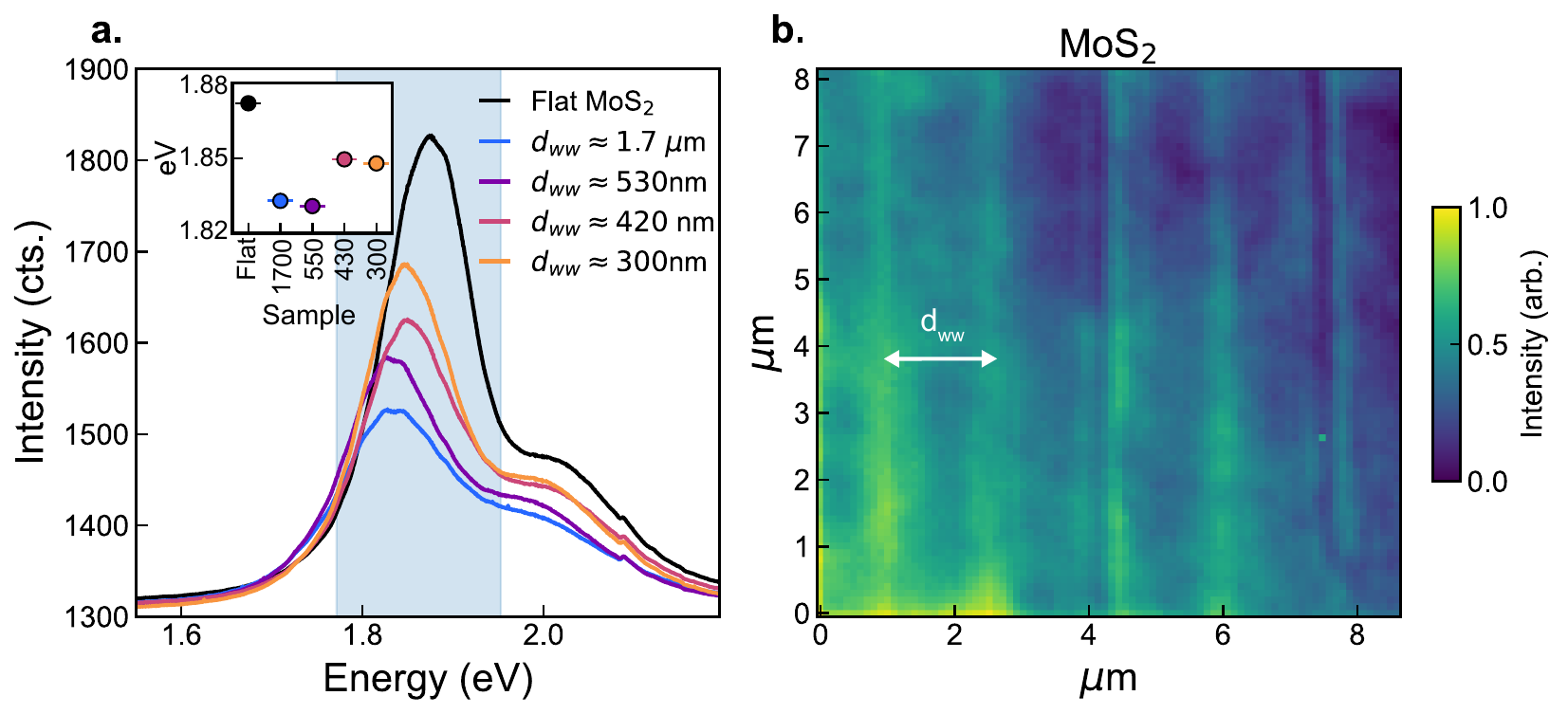}
   \caption{(a) Photoluminescence spectra of Flat and NW-MoS$_2$ of various wrinkle densities. $\text{d}_\text{ww}$ : median wrinkle spacing. Spectra were averaged over 5 × 5 $\mu$m$^2$ regions to account for local heterogeneity.  Inset : PL peak position of flat monolayer and monolayers with different wrinkle spacings.  (b) Integrated PL intensity over blue shaded energy window in (a) for NW-MoS$_2$ with $\text{d}_\text{ww} \approx 1.7\mu$m. The higher intensity regions appear at the same periodicity  $\text{d}_\text{ww}$ as the wrinkled regions from AFM images. The PL measurements were conducted using a 532 nm excitation laser with a spot size of $\approx$ 420 nm.} 
   \label{fig:2}
\end{figure}

Bubbles in monolayer materials induce local variations of strain and contact, which in turn modulate their optical and thermal properties. \cite{zhu_thermal_2015, ding_manipulating_2015, shafique_strain_2017, liu_reduction_2025}.To illustrate such effect in the periodically deformed monolayers, we performed photoluminescence (PL) characterization on flat and nano-wrinkled (NW) MoS$_2$ monolayers with different wrinkle periodicity. The results are shown in Figure \ref{fig:2}a. Compared with the flat monolayer, the nano-wrinkled monolayers exhibit PL peak shift by $\approx$ 22 to 41 meV to lower energy, accompanied by a notable reduction in PL intensity. The extent of intensity quenching correlates with the magnitude of the redshift, suggesting an overall effect of tensile strain averaged across the majority of the monolayer. When MoS$_2$ monolayers are placed under biaxial tensile strain, the conduction band minimum at the $K$ point is lowered while the valence band at the $\Gamma$ point is raised, effectively narrowing the band gap \cite{conley_bandgap_2013}. At sufficiently high strain levels, MoS$_2$ monolayer turns into an indirect gap semiconductor, suppressing radiative recombination and quenching the PL emission. \cite{chang_orbital_2013, yun_thickness_2012}

To further explore spatial variations across individual wrinkles, we performed spatially resolved PL mapping on MoS$_2$ monolayers with wrinkle periodicities of $\approx 1.7\mu$m, much larger than the laser spot size ($\approx$ 420 nm). The PL spectrum measured at each spot predominantly reflects the local geometry, strain, and interaction with the substrate within the illuminated area. As shown in Figure \ref{fig:2}b, the PL intensity exhibits periodic modulation at intervals aligned with the wrinkle spacing. Notably, the apices of the wrinkles where the monolayer is suspended display enhanced PL emission compared to the adjacent substrate-supported regions. This enhancement in the freestanding areas is attributed to reduced dielectric screening, suppressed substrate-induced scattering, reduced nonradiative recombination and longer carrier lifetimes. Moreover, the nonuniform strain distribution in the wrinkles are reported to modulate the optical bandgap locally, generating potential wells that funnel photogenerated excitons toward the lower-bandgap regions, where they preferentially recombine \cite{castellanos-gomez_local_2013}. 

To further evaluate the strain distribution, we performed Raman spectroscopy on the NW-MoS$_2$ monolayers. As shown in Figure 3a and 3b, all NW-MoS$_2$ samples exhibit a shift of both the $E'$ and $A_1'$ phonon modes to lower frequencies compared with flat monolayer MoS$_2$. This softening of vibrational modes indicates the presence of tensile strain, as reported in previous Raman studies of monolayer MoS$_2$ under uniaxial and biaxial strain\cite{castellanos-gomez_local_2013, tan_raman_2018, rice_raman-scattering_2013}. The observation of tensile strain is also consistent with our PL measurements. 

\begin{figure}[H]
   \centering
   \includegraphics[width=0.8\linewidth]{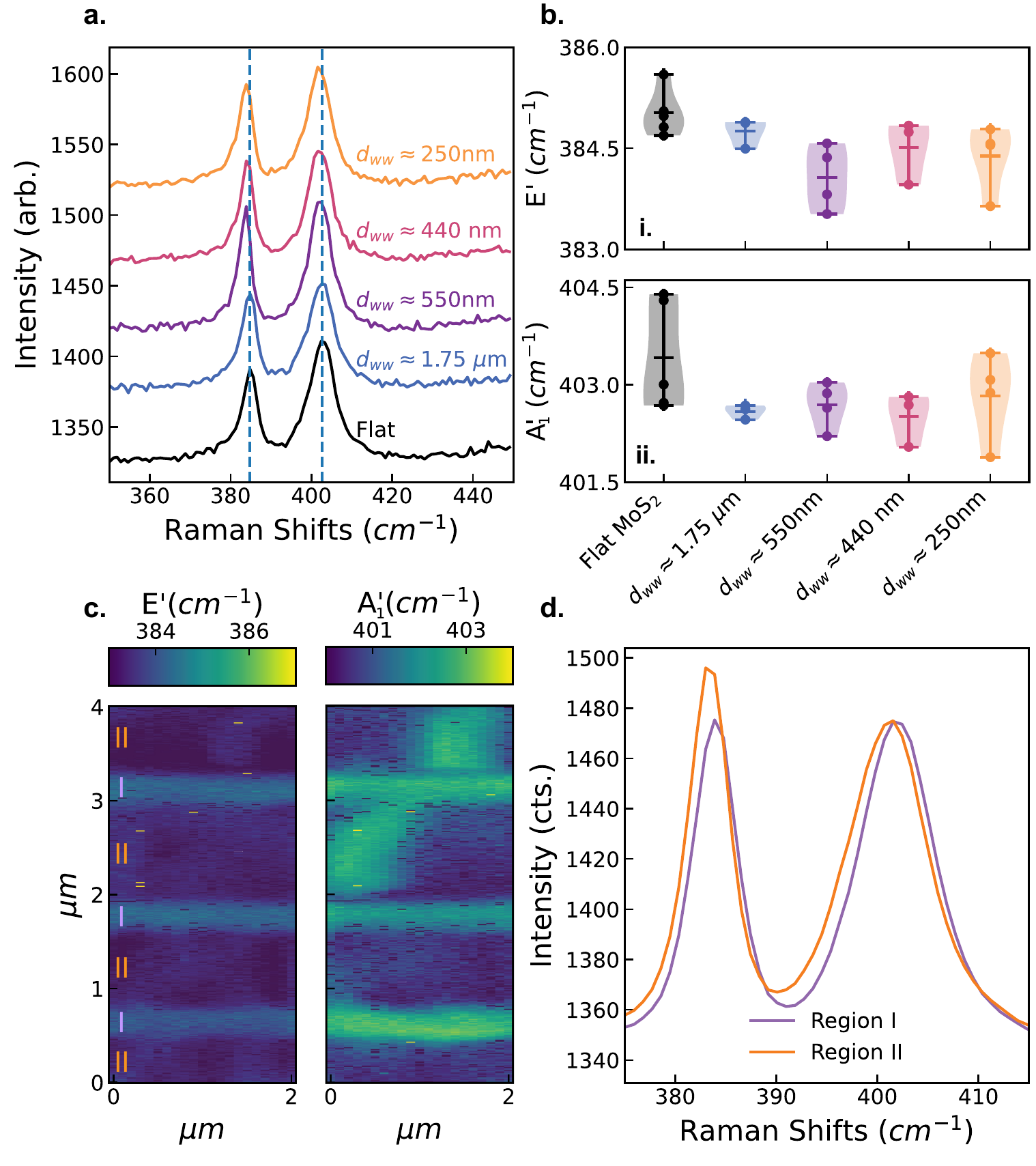}
   \caption{(a) Representative Raman spectra of flat and NW-MoS$_2$ samples. Vertical dashed lines indicate the peak positions of the flat monolayer for reference. Spectra were taken with a spot size of $\approx 1 \mu$m and averaged over 4 $\times$ 4 $\mu$m$^2$ regions across samples with varying wrinkle densities. (b) Extracted peak positions for the $E'$ (i) and $A_1'$ (ii) modes as a function of wrinkle density, showing systematic redshifts in NW-MoS$_2$ relative to the flat monolayer. Circles correspond to measurements on different regions on sample. Horizontal bars correspond to minimum, mean, and maximum of measurements. Shaded regions correspond to the estimated probability density function of the measurements using a gaussian kernel density estimation. (c) Spatial maps of Raman peak positions for the $E'$ and $A_1'$ modes in a NW-MoS$_2$ sample with a median wrinkle spacing of $\approx 1.7\mu$m, measured with a spot size of $\approx$ 420 nm. (d) Averaged Raman spectra from blue-shifted (Region I) and red-shifted (Region II) areas.}
   \label{fig:3}
\end{figure}

For local interrogation of both wrinkled and flat regions, we demonstrate spatially resolved Raman mapping, as shown in Figure 3c. The positions of the $E'$ and $A_1'$ Raman peaks are spatially modulated at intervals consistent with the wrinkle spacing. These alternating phonon frequency shifts to lower and higher frequencies correspond to the strained and less-strained areas of the monolayer, respectively. Spectra averaged on these two areas (Figure \ref{fig:3}d) demonstrate a relative shift of $\approx 0.58$ cm$^{-1}$ of the $E'$ mode and $\approx 0.85$ cm$^{-1}$ of the $A_1'$ mode, confirming the presence of spatially varying strain by 1.4-3$\%$ \cite{lloyd_band_2016}. Correlated Raman and PL maps from the same region show that areas with Raman modes shifting to lower frequencies indicating enhanced tensile strain coincide with areas of reduced PL intensity. Interestingly, these regions correspond not to the wrinkle apex but to the flat regions between wrinkles that remain in contact with the substrate. This suggests that the highest tensile strain occurs in the monolayer segments that are pushed downward into the lower contours of the mold during transfer. As the monolayer conforms to the underlying topography, these bottom-contacted regions experience the greatest in-plane stretching, resulting in stronger phonon softening and a shift toward indirect bandgap characteristics. 

\section{Thermal Conductivity Measurements}

\begin{figure}[H]
   \centering
   \includegraphics[width=0.9\linewidth]{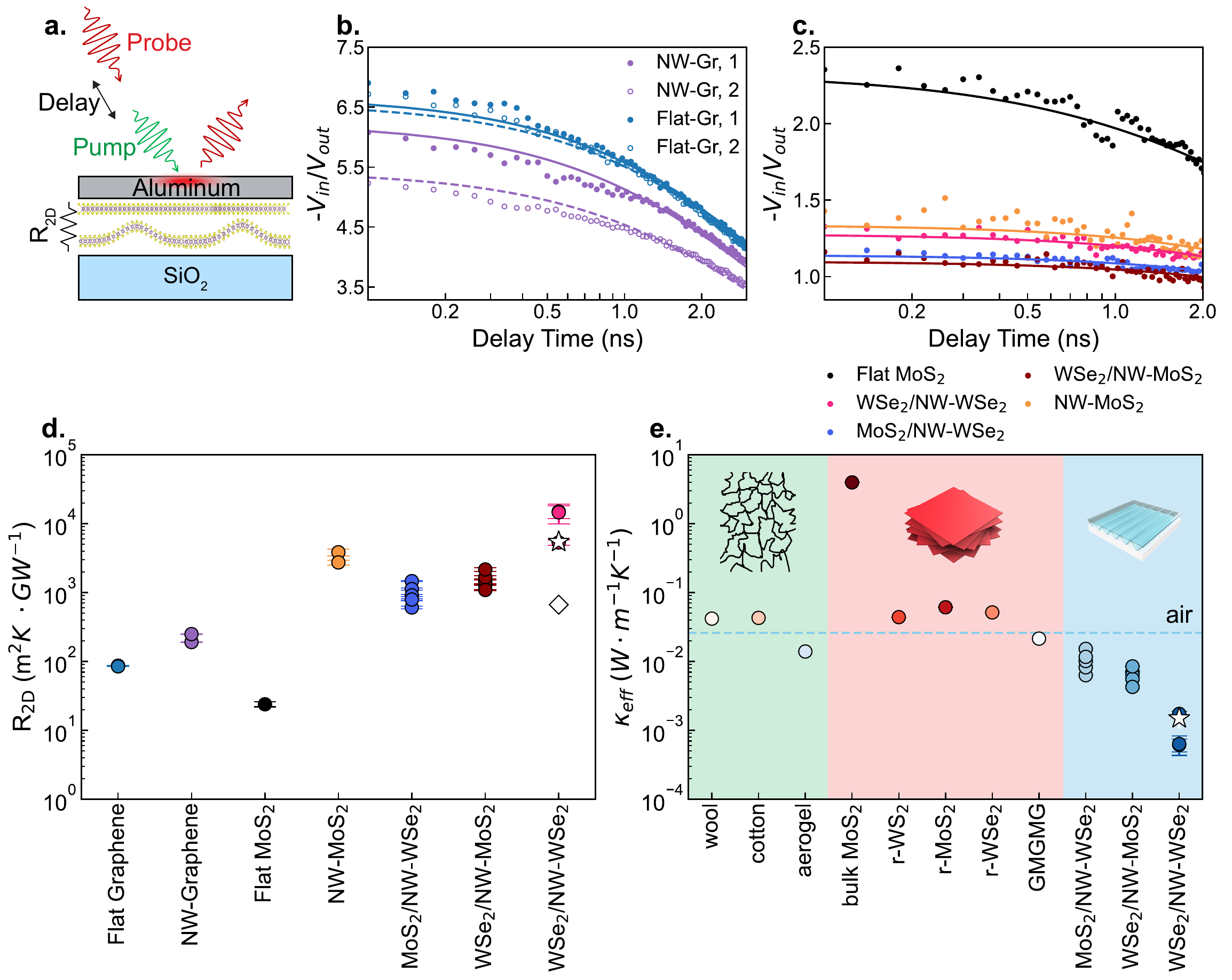}
   \caption{(a) Schematic of the TDTR  measurement setup. (b) Representative TDTR signals for flat and NW graphene samples.(c) Representative TDTR signals for flat TMDC monolayers, NW TMDC monolayers, and monolayer bubble wrap structures formed by stacking flat and NW monolayers. (d) Extracted thermal resistance values R$_\text{2D}$ of samples in (b) and (c). White star: calculated overall thermal resistance of WSe$_2$ bubble wrap  structure. White diamond: calculated combined interfacial thermal resistances of top and bottom interfaces in the WSe$_2$ nano-bubble wrap structure. (e) Comparision of the effective thermal conductivity of nano-bubble wrap architectures with conventional low thermal conductivity materials, including aerogels.\cite{Engineeringtoolbox_2025}\cite{lu_thermal_1992} \cite{chiritescu_ultralow_2007}\cite{kim_extremely_2021}\cite{sood_engineering_2021}} 
   \label{fig:4}
\end{figure}

To evaluate the thermal transport properties through periodically deformed monolayers and bilayer bubble-wrap structures, we employed time-domain thermoreflectance (TDTR) measurements (Figure 4a). TDTR is a well-established pump-probe technique widely used to characterize thermal transport across thin-films and interfaces\cite{jiang_tutorial_2018}. Details of the experimental setup used in this work can be found in the supplementary. In brief, a 60nm thick aluminum transducer layer was deposited onto the samples using electron-beam deposition. Al was selected for its high thermoreflectance coefficient $dR/dT \approx 10^{-4}$ $K^{-1}$ at 800nm, rapid heat dissipation, and strong optical absorption \cite{jiang_tutorial_2018}. The pump beam, modulated at frequencies $f_\text{mod}$ = 3 MHz, was focused on the Al surface with root-mean-square $(1/e^2)$ spot diameter of $\approx 3-4 \mu$m. The large spot size makes these measurements sensitive to out-of-plane thermal transport, and effectively averages over many wrinkles. This provides representative values of out-of-plane thermal conductivity and interfacial thermal resistance for the nano-bubble wrapped materials.

We used TDTR to characterize thermal transport in flat and NW MoS$_2$ and graphene monolayers, as well as in heterostructured “bubble wrap” samples consisting of a flat monolayer overlying a NW monolayer, thus forming a structure reminiscent of conventional bubble wrap. To extract the effective thermal properties from the multilayer stack configuration (Al/2D material/SiO$_2$/Si), we adopted the reported thermal conductivities of Al, SiO$_2$, and Si, along with the volumetric specific heats of Al, SiO$_2$, the 2D materials, and Si \cite{goni_enhanced_2018, regner_broadband_2013, zhang_thermal_2023, linstrom_nist_2001, bolgar_thermodynamic_1990}. For the periodically deformed monolayer and bilayer samples, we calculate the volumetric specific heat of the mixture of the constituent materials and air. Owing to the atomically thin nature of monolayer 2D materials, prior studies have shown that long-wavelength phonons can traverse the metal transducer layer and couple directly to the underlying SiO$_2$ substrate with minimal scattering, particularly in flat MoS$_2$/SiO$_2$ systems. Consequently, TDTR measurements on 2D material samples are typically modeled using a single fitting parameter: the thermal resistance between the transducer layer and the SiO$_2$ substrate, R$_\text{2D}$, or conversely, the effective out-of-plane thermal boundary conductance (G$_\text{eff}$) \cite{sood_engineering_2021, goni_enhanced_2018, koh_heat_2010} . We adapt a similar modeling approach. Details of the fitting procedure, model assumptions, and parameters used for fitting are provided in the Supplementary Information.

Representative TDTR measurements for flat and NW monolayers of MoS$_2$ and graphene, along with the corresponding extracted interfacial thermal resistance, R$_\text{2D}$, are presented in Figure 4b-c. Clear differences are observed between flat and NW samples in both the TDTR signal intensity and decay rate, indicating that R$_\text{2D}$ is substantially enhanced in the NW monolayers. Figure 4d shows the extracted R$_\text{2D}$ for the measured samples. For flat monolayer samples, we extract R$_\text{2D}$ values of approximately 24 $\text{m}^{2}\, \text{K} \cdot \text{GW}^{-1}$ for MoS$_2$ and 85 $\text{m}^{2}\, \text{K} \cdot \text{GW}^{-1}$ for graphene, consistent with previously reported values in the literature \cite{goni_enhanced_2018, sood_engineering_2021}. In stark contrast, NW counterparts show a pronounced enhancement in interfacial thermal resistance. Specifically, two NW-graphene samples yield R$_\text{2D}$ values of 190 and 250 $\text{m}^{2}\, \text{K} \cdot \text{GW}^{-1}$ corresponding to a 2 to 3-fold increase relative to flat graphene. The enhancement is even more dramatic for NW-MoS$_2$, with R$_\text{2D}$ values as high as 2.7-3.8 $\times 10^3$ $\text{m}^{2}\, \text{K} \cdot \text{GW}^{-1}$ – up to a two orders of magnitude increase compared to its flat counterpart.

Finally, we demonstrate that stacking a flat monolayer atop a nano-wrinkled monolayer—forming a ``monolayer bubble wrap" structure—yields an even more dramatic suppression of interfacial heat transport. In this configuration, the R$_\text{2D}$ increases by nearly another order of magnitude to values as high as $10^4$ $\text{m}^{2}\, \text{K} \, \cdot \, \text{GW}^{-1}$, representing a nearly three-orders-of-magnitude enhancement compared to pristine, conformal monolayer interfaces. Compared with previous reports on thermally resistive vdW interfaces, such as graphene/MoS$_2$ multilayer heterostructures\cite{sood_engineering_2021} or rotationally mismatched MoS$_2$, WS$_2$, or WSe$_2$, multilayer stacks\cite{kim_extremely_2021, chiritescu_ultralow_2007}, where heat transport is hindered by phonon spectral mismatch, poor adhesion, or interfacial disorder our measured R$_\text{2D}$ is more than one order of magnitude higher. This ultralow thermal coupling places our system among the most thermally resistive vdW heterostructures reported to date.

From the measured thermal resistance R$_\text{2D}$, we estimate \emph{effective} thermal conductivities, shown in Figure \ref{fig:4}d. The thermal conductivities of nano-bubble wraps are remarkably low, reaching $\approx$ 0.001 W $\cdot$ m$^{-1}$ $K^{-1}$ under ambient conditions, which is an order of magnitude lower than that of air and even silica aerogels\cite{lu_thermal_1992}, the most thermally insulating materials known. This places our structures well within the deep-Knudsen regime and highlights a new nanoscale design principle for achieving extreme thermal insulation beyond the limits of conventional porous materials.

The changes in resistance are substantially greater than what would be expected from strain effects alone. The averaged tensile strain in the flat contact regions is approximately 1-3$\%$ from Raman measurements. Such strain levels, however, would only result in a modest enhancement of 2-5X in thermal resistance for TMDC monolayers, as revealed by previous studies on strain-dependent interfacial thermal transport\cite{liu_reduction_2025}. In general, out-of-plane thermal boundary resistance can be influenced by three primary factors: phonon dispersion, temperature, and the interfacial coupling between materials\cite{monachon_thermal_2016}. The significantly larger reduction observed in the nano-wrinkled monolayers likely arises from the combined effects of two main phenomena. First, the presence of periodic nanoscale air gaps between the 2D material and the underlying substrate serve as highly resistive thermal barriers. Second, the presence of periodic topographic modulations with characteristic dimensions on the order of the phonon mean free path can alter phonon dispersions and weaken interfacial coupling, further impeding phonon-mediated cross-plane heat flow. Together, these effects result in a pronounced reduction in thermal conductance through the nano-bubble wrap materials, underscoring the potential of nano-structured deformation as an effective strategy for tuning interfacial heat transport in ultrathin material systems.

To understand the origin of the extremely high thermal resistance observed in the WSe$_2$ nano–bubble wrap structure, we model the system using established frameworks developed for porous material such as aerogels \cite{fu_critical_2022}. In this approach, the effective thermal conductivity is expressed as $\kappa = (1 - \Pi)\kappa_s + \Pi\kappa_g$, where $\Pi$ is the porosity, $\kappa_s$ is the intrinsic thermal conductivity of the solid skeleton, and $\kappa_g$ is the thermal conductivity of the confined gas. Due to the atomically thin nature of the monolayer walls, the porosity of the structure is exceptionally high. Based on the thickness of an individual layer of WSe$_2$ of $\approx$ 3.3 \AA \cite{SCHUTTE1987207} and a periodicity of 300nm, we estimate $\Pi \approx 99.78\%$. At such high porosity, a substantial fraction of heat is expected to be carried through the gas phase. However, in conventional aerogels, increasing porosity typically results in larger pore sizes, which enhances both gas-phase conduction and radiative heat transfer, ultimately limiting further reduction in thermal conductivity. In contrast, our nano–bubble wrap  architecture combines extremely high porosity with nanoscale gas confinement deep in the Knudsen regime, where molecular collisions with the pore walls dominate over intermolecular collisions \cite{fu_critical_2022}, offering a new pathway to achieving exceptionally low thermal conductivity even under ambient conditions.  

We estimate the thermal conductivity of confined air using the Kaganer model, which describes heat transport between two parallel plates separated by a nanoscale air gap\cite{kaganer1969thermal}. Details of the model implementation are provided in the Supplementary Information. The effective thermal conductivity of the trapped air, $\kappa_g$, is calculated by statistically averaging over the air gap distribution in the nano-bubble wrap  structures (Supplementary Figure S4), which is centered around $\approx 8.61$ nm. Using this approach, we obtain a value of $\kappa_g \approx$ 0.001 W$\cdot$m$^{-1} $K$^{-1}$. 

The solid component of the structure consists of the WSe$_2$ monolayer walls with a height of $\approx 8.61$ nm. At this dimension, which is well below the phonon mean free path, thermal transport is significantly suppressed. Prior theoretical studies estimate that the thermal conductivity of WSe$_2$ nanoribbons at this scale is reduced by approximately 95\% compared to 2D monolayers, yielding an estimated value of $\kappa_s \approx 0.23$W$\cdot$m$^{-1} $K$^{-1}$ for 10nm wide WSe$_2$ monolayer nanoribbons.\cite{zhou_first-principles_2015} Given the porosity $\Pi \approx 99.78\%$, the corresponding contribution from the monolayer framework is $(1 - \Pi)\kappa_s \approx 0.0005$~W$\cdot$m$^{-1}\, $K$^{-1}$.

Combining the contributions from the nanometer-scale air gap and the narrow monolayer walls, the total thermal conductance of the porous structure is estimated to be $\approx 0.0015 \text{W} \cdot \text{m}^{-1}\, \text{K}^{-1}$. Additional resistance arises from the thermal boundary conductance at the bottom WSe$_2$/SiO$_2$ interface \cite{hunter_interfacial} and the weakly coupled top WSe$_2$/WSe$_2$ interface \cite{chiritescu_ultralow_2007}. As detailed in the full thermal model in the Supplementary Information and highlighted by the white diamond in Figure 4d, these interfacial contributions are relatively minor compared to the dominant resistance originating from the porous region. 

Taking all contributions above into account, the total thermal resistance of the WSe$_2$ nano–bubble wrap structure is estimated to be $\text{R} \approx$ 5447.2 $\pm$ 914.9 m$^2\cdot$K$\cdot$GW$^{-1}$, as marked by the star in Fig. \ref{fig:4}d. This predicted value is in excellent agreement with our experimental measurements. While the current model captures the dominant mechanisms governing thermal resistance, further refinements could enhance its accuracy. These include incorporating the thermal boundary conductance at the air–monolayer interface, as well as the effects of local strain, atomic-scale disorder, and curvature-induced variations within the bubble regions—all of which are expected to further reduce thermal conductivity. Additionally, improved treatment of the top surface morphology and a more rigorous description of phonon transport in confined monolayer ribbons would provide a more complete picture of heat transfer in these systems. 

From an experimental perspective, several strategies may be pursued to further enhance thermal insulation. These include the use of disordered polycrystalline monolayers to increase phonon scattering, replacing air with lower-thermal-conductivity gases such as krypton or xenon to suppress gas-phase conduction, and vertically stacking multiple bubble wrap  layers to create a three-dimensional insulating architecture. The latter approach is compatible with existing fabrication techniques and can be readily implemented using automated robotic stacking systems \cite{mannix_robotic_2022}, offering a practical route toward scalable materials with ultralow thermal conductivity.

\section{Discussion \& Conclusion}

We demonstrate that nano–bubble wrap structures with precisely engineered nanoscale morphology offer a powerful approach to modulate the thermal properties of monolayer materials, with pronounced effects on out-of-plane thermal transport. We use TDTR to reveal a consistent reduction on thermal transport in NW-monolayers compared to their flat counterparts. The nano-bubbles formed in these structures, with dimensions well below the mean free path of air, place the system deep within the Knudsen regime, achieving thermal conductivities nearly an order of magnitude lower than that of air. Beyond gas confinement, the periodic strain and nanoscale deformation in monolayers alter the local phonon frequency and dispersion, which in turn modulates phonon transmission across the vdW interface. Further, the reduced dimensionalities of monolayer gaps comparable with the phonon mean free path enhances phonon scattering, thus lowering thermal conductivity \cite{zhu_thermal_2015, chang_orbital_2013}. The weak interfacial coupling at the vdW boundaries between bubble wrap  layers and with the underlying substrate also significantly reduces the thermal transport, contributing to the overall ultralow thermal conductivity observed.

Unlike previous studies that have focused on localized strain in isolated bubbles or wrinkles,\cite{castellanos-gomez_local_2013, darlington_imaging_2020} or employed external substrates and bending setups to induce average strain\cite{mennel_second_2019}, our method  intrinsically generates large-area, spatially modulated uniaxial strain intrinsic to the monolayer with flat contact area with a standard substrate. This scalable fabrication technique can provide a versatile platform for creating fundamental strain-induced phenomena in two-dimensional materials. This can also lead to integration of these materials into electronic devices without the need for complex external strain setups. Beyond strain engineering, these bubble wrap structures offer exciting opportunities for tuning interfacial adhesion and thermal mismatch at the nanoscale, enabling spatial control over phonon coupling and mechanical contact. Such localized modulation of nanoscale energy transport may open new avenues for directing heat flow, with implications for advanced thermal management applications.

\begin{acknowledgement}

Preparation of nano-bubble wraps at Stanford is
supported by the Defense Advanced Research Projects Agency (DARPA) under Agreement No. HR00112390108. Data analysis for TDTR in this work was  supported by the U.S. Department of Energy, Office of Science, for support of  microelectronics research, under contract number DE-AC02-05CH11231. Raman and PL micro-spectroscopy and TDTR measurements at the Molecular Foundry were supported by the Office of Science, Office of Basic Energy Sciences, of the U.S. Department of Energy under Contract No. DE-AC02-05CH11231. A.C.J. acknowledges support from the Stanford TomKat Institute Graduate Translational Fellowship. A.C.J. also acknowledges support from the the U.S. Department of Energy, Office of Science, Office of Workforce Development for Teachers and Scientists, Office of Science Graduate Student Research (SCGSR) program. The SCGSR program is administered by the Oak Ridge Institute for Science and Education for the DOE under contract number DE‐SC0014664. Frank Ogletree constructred the TDTR system, and he and Sean Lubner trained S.W. on its use.

\end{acknowledgement}

\subsection{Author contributions}
F.L. conceived the project. A.R. and F.L. supervised the project. A.C.J. prepared the nano-bubble wrap materials. A.C.J performed the Raman and Photoluminescence measurements and analyzed the data with help from F.L. and A.R. A.C.J performed TDTR measurements and analyzed the data with help from S.W. and A.R. All authors discussed the results and contributed to writing the manuscript.


\clearpage
\bibliography{bibliography}


\end{document}